# Bright radio emission from an ultraluminous stellar-mass microquasar in M31

Authors: Matthew J. Middleton[1,2], James C.A. Miller-Jones[3], Sera Markoff[2], Rob Fender[4], Martin Henze[5], Natasha Hurley-Walker[3], Anna M.M. Scaife[4], Timothy P. Roberts[1], Dominic Walton[6,7], John Carpenter[7], Jean-Pierre Macquart[3,8], Geoffrey C. Bower[9], Mark Gurwell[10], Wolfgang Pietsch[5], Frank Haberl[5], Jonathan Harris[1], Michael Daniel[1], Junayd Miah[1], Chris Done[1], John Morgan[3], Hugh Dickinson[11], Phil Charles[4,12], Vadim Burwitz[5], Massimo Della Valle[13], Michael Freyberg[5], Jochen Greiner[5], Margarita Hernanz[14], Dieter H. Hartmann[15], Despina Hatzidimitriou[16], Arno Riffeser[17], Gloria Sala[18], Stella Seitz[17], Pablo Reig[19], Arne Rau[5], Marina Orio[20], David Titterington[21], Keith Grainge[21]

Affiliations:
1. Physics Department, University of Durham, Durham, County Durham, DH1 3LE, UK
2. Astronomical Institute Anton Pannekoek, Science Park 904, 1098 XH, Amsterdam, Netherlands
3. International Centre for Radio Astronomy Research, Curtin University, GPO Box U1987, Perth, WA 6845, Australia
4. School of Physics & Astronomy, University of Southampton, Southampton, Hampshire, SO17 1BJ, UK
5. Max-Planck-Institut für extraterrestrische Physik, Giessenbachstrasse 85748, Garching, Germany
6. Institute of Astronomy, Cambridge University, Madingley Road, Cambridge, CB3 0HA, UK
7. Astronomy Department, California Institute of Technology, MC 249-17 1200 East California Blvd, Pasadena, California, 91125, US
8. ARC Centre of Excellence for All-Sky Astrophysics (CAASTRO), Curtin University, GPO Box U1987, Perth WA 6845, Australia
9. Astronomy Department, B-20 Hearst Field Annex # 3411, University of California at Berkeley, Berkeley, California 94720-3411, US
10. Harvard-Smithsonian Center for Astrophysics, 60 Garden Street, Cambridge, Massachusetts 02138, US
11. Stockholm University, Oskar Klein Centre, AlbaNova, SE-106 91, Stockholm, Sweden
12. Department of Astronomy, University of Cape Town, Private Bag X3, Rondebosch 7701, Republic of South Africa
13. Osservatorio Astronomico di Capodimonte, INAF, Salita Moiariello 16, 80131, Napoli, Italy
14. Institute of Space Sciences (CSIC-IEEC), Campus UAB, Fac. Ciències, Torre C5 parell 2. 08193 Bellaterra, Barcelona, Spain
15. Physics & Astronomy Department, 118 Kinard Laboratory, Clemson University, Clemson, South Carolina 29631-0978, US
16. Department of Astrophysics, Astronomy and Mechanics, Faculty of Physics, National and Kapodistrian University of Athens, Panepistimiopolis, GR 157 84 Zografou, Athens, Greece
17. University Observatory Munich, Ludwig-Maximilians-Universität, Scheinerstr. 1, D-81679 München, Germany

18. Department of Physics and Nuclear Engineering, EUETIB (UPC-IEEC), c/ Comte d'Urgell 187, E-08036 Barcelona, Spain
19. Foundation for Research & Technology-Hellas, Nikolaou Plastira 100, Vassilika Vouton GR - 711 10, Heraklion, Crete, Greece
20. Osservatorio Astronomico di Padova, Vicolo Osservatorio 5 – 35122, Padova, Italy
21. Astrophysics Group, Cavendish Laboratory, Cambridge University, JJ Thomson Avenue, Cambridge, CB3 0HE, UK

**A subset of ultraluminous X-ray sources (those with luminosities < $10^{40}$ erg s$^{-1}$)[1] are thought to be powered by the accretion of gas onto black holes with masses of ~5-20 $M_\odot$, probably via an accretion disc[2,3]. The X-ray and radio emission are coupled in such Galactic sources, with the radio emission originating in a relativistic jet thought to be launched from the innermost regions near the black hole[4,5], with the most powerful emission occurring when the rate of infalling matter approaches a theoretical maximum (the Eddington limit). Only four such maximal sources are known in the Milky Way[6], and the absorption of soft X-rays in the interstellar medium precludes determining the causal sequence of events that leads to the ejection of the jet. Here we report radio and X-ray observations of a bright new X-ray source whose peak luminosity can exceed $10^{39}$ erg s$^{-1}$ in the nearby galaxy, M31. The radio luminosity is extremely high and shows variability on a timescale of tens of minutes, arguing that the source is highly compact and powered by accretion close to the Eddington limit onto a stellar mass black hole. Continued radio and X-ray monitoring of such sources should reveal the causal relationship between the accretion flow and the powerful jet emission.**

*XMM-Newton* first detected XMMU J004243.6+412519 on January 15th 2012[7] at an X-ray luminosity of $2 \times 10^{38}$ erg s$^{-1}$ (for a distance to M31 of 0.78 Mpc[8]), with an X-ray spectrum that could be fully described by a hard power-law, characteristic of sub-Eddington accretion (mass accretion rates < 70% Eddington[9,10]). The source then rose to > $1 \times 10^{39}$ erg s$^{-1}$ in two subsequent detections, fulfilling the traditional definition of an ultraluminous X-ray source (ULX - while other definitions exist, the term ULX is numerical rather than physically motivated)[11] and significantly above the cut-off luminosity of the X-ray luminosity function of M31[12], which shows no sources more luminous than $2\times10^{38}$ erg s$^{-1}$.

At the peak luminosity of $1.26 \pm 0.01 \times 10^{39}$ erg s$^{-1}$, the X-ray spectrum appeared similar to that of Galactic black hole X-ray binaries (BHXRBs) at mass accretion rates close to or above the Eddington limit[13]. In such cases, the emission is dominated by an optically thick accretion disc[2], whose spectrum may appear broadened as the accretion process is no longer radiatively efficient[14]. This can be accompanied by a second, weaker, thermal component at higher energies, possibly due to Compton up-scattering of disc photons in a wind or photosphere. These two components are also required to fit high-quality spectra of nearby, 'low luminosity' ULXs[3,15], implying similar accretion processes are taking place. Although the intrinsic emission can potentially be amplified through geometrical beaming[16] this is thought to be important only for ULXs above $10^{40}$ erg s$^{-1}$.

Following the X-ray outburst, the source was monitored by the *Swift* and *Chandra* missions for over eight weeks (see Figure 1 and Supplementary table 1), until the source was no longer observable due to the Sun angle. The X-ray luminosity decreased slightly to $\sim 1\times 10^{39}$ erg s$^{-1}$, with the spectrum evolving to be fully described by emission from a standard accretion disc[2] with a peak temperature characteristic of high mass accretion rate Galactic BHXRBs[9]. Both the timescale for the X-ray spectral variations and the temperature of the disc component, rule out an interpretation as a background active galaxy.

A second, subsequent period of X-ray monitoring showed the source to remain disc-dominated whilst decaying to a luminosity of $\sim 7\times 10^{37}$ erg s$^{-1}$. The correlation of luminosity with disc temperature as $L \propto T^4$ is a well established observational tracer of sub-Eddington accretion. We find that the best-fitting spectral parameters (see Supplementary Table 2) deviated strongly from this correlation when the source was brightest, with the temperature having decreased with luminosity (Supplementary Figure 2). Such behaviour is well documented in ULXs[17], strengthening our identification of this object as a - albeit low-luminosity - member of this class.

The X-ray detection of this nearby, bright source motivated a series of radio observations by the Karl G. Jansky Very Large Array (VLA), Very Long Baseline Array (VLBA) and Arcminute Microkelvin Imager Large Array (AMI-LA). The radio light curves are shown in Figure 2, with the observational details provided in the SI. The source was initially detected by the VLA at a highly significant level (> 40σ), with a 4-5σ AMI-LA detection the following day. The spectral index (defined by $S_\nu \propto \nu^\alpha$) measured by the VLA was slightly inverted, at α = 0.27 ± 0.16.

The AMI-LA monitoring demonstrated that the radio emission was variable on timescales of days, as confirmed by a detection of strongly reduced emission by the VLA (Figure 2) and a non-detection by the Combined Array for Research in Millimeter-wave Astronomy (CARMA). Previous, well-characterised, radio-luminous ULXs have been identified with emission from an optically thin, diffuse nebula believed to be powered by the bright accretion flow[18]. The radio variability from XMMU J004243.6+412519 demonstrates that the emission cannot be nebular in origin. Furthermore, the unresolved VLBA detection, shown in Figure 2, constrains the size of the emitting region to be < 1500 AU. This size limit, together with the day-timescale variability, confirms that the radio emission must originate in a relatively compact, relativistically outflowing region with a brightness temperature > $6\times 10^6$ K.

BHXRBs accreting at sub-Eddington rates produce powerful, persistent radio jets, with the radio and X-ray emission strongly correlated, and linked to black hole mass via the "fundamental plane of black hole activity[19,20]". However, the thermal, disc-dominated X-ray spectrum following the peak of the outburst implies accretion at a relatively high Eddington fraction[10] (3-100%). The fundamental plane relation does not hold for sources accreting at such high mass accretion rates and so cannot be used to constrain the black hole mass. At these high accretion rates, the radio emission from BHXRBs is instead observed to undergo flaring periods, associated with transitions from hard to soft X-ray spectral states[5]. Indeed, scaling the radio and X-ray flux from this source to be within our Galaxy would give fluxes similar to soft, bright, radio-flaring BHXRBs such as GRS 1915+105 and Cygnus X-3[21]. The observed X-ray and radio emission of this source is therefore fully consistent with analogous behaviour and the radio emission is thus likely associated with high bulk Lorentz factor ejections accelerating particles within a shocked plasma[22].

The AMI-LA light curve (Figure 2) suggests multiple ejection events, although the observed inverted, optically-thick spectrum is at odds with the optically-thin emission expected from the later stages of an expanding synchrotron-emitting plasma. Either the VLA observed the early, self-absorbed stage of a flare, or there is additional, free-free absorbing material in the environment of the source.

A search for shorter-timescale fluctuations in the VLA data revealed significant variability in the first epoch, on a characteristic timescale of several minutes (Figure 2). Regardless of whether we attribute this to intrinsic variability or to scintillation (see SI), this timescale implies a source size of only ~5 AU (a few microarcseconds at the distance of M31), constraining the emitting region to be highly compact.

There are clear similarities between the behaviour of XMMU J004243.6+412519 and the few 'super-Eddington' Galactic BHXRBs. Of these, the canonical microquasar, GRS 1915+105 is the only one with mass accretion rates *regularly* at or in excess of $1\times10^{39}$ erg/s, and would sometimes appear as a ULX to an extragalactic observer[23]. Multi-wavelength studies of the disc-jet coupling in GRS 1915+105[24] have identified both 'plateau' states with steady jet emission at lower mass accretion rates, and 'flaring' states with rapid radio oscillations at accretion rates around the Eddington limit, accompanied by an extremely soft X-ray spectrum. The latter, minute-timescale[25] radio flaring has been observed to occur immediately following a major radio flare, accompanied by X-ray brightness changes that are dominated by variability above 3 keV. Should the short-timescale radio variability we observe be intrinsic (see SI), it could be an analogue of this behaviour, although we would not expect to detect the X-ray variability in our observations, since the flux above 3 keV is extremely low due to the greater source distance. Although GRS 1915+105 shows clear analogous behaviour, when scaled to a similar distance, our source is approximately an order of magnitude more luminous in the radio band, with a variability brightness temperature for the flares of $\sim 7 \times 10^{10}$ K (see SI). Such a discrepancy can be readily explained by differences in the inclination angle of the jets to our line-of-sight; GRS1915+105 is highly-inclined and thus Doppler-deboosted, whereas we may be looking closer to the jet axis in XMMU J004243.6+412519. This physical picture is supported by comparison to the Galactic BHXRBs, Cygnus X-3 and V4641 Sgr, both of which have reached similarly high radio luminosities and probably have low angles to the line-of-sight[26,27]. Our source seems particularly well matched to V4641 Sgr, which reached ~0.5 Jy in its 1999 outburst[28], and has also exhibited similarly high brightness temperatures.

The coupled radio and X-ray behaviour in this source is fully consistent with a picture of a BHXRB rising above the Eddington limit in outburst then dimming to a bright sub-Eddington state, with radio flaring analogous to that seen in GRS 1915+105 and other BHXRBs[5]. We can therefore make a secure identification of the compact object in this source as a stellar mass (< 70 $M_\odot$) black hole.

As the source remains in a disc-dominated state down to ~$7\times10^{37}$ erg s$^{-1}$, and the lower limit for such a state is ~3% of Eddington[10], this constrains the black hole mass to < 17 $M_\odot$. However, as the joint radio/X-ray behaviour implies Eddington-rate accretion[5] at the peak luminosity of ~$1.3\times 10^{39}$ erg s$^{-1}$, we favour a more conservative mass estimate of ~10 $M_\odot$.

We can also constrain the nature of the companion star from archival optical data[29]. The field does not contain a source within 3 arcsec (where the positional accuracy from the VLBA observations is extremely high - see SI) down to V ~ 25 and B ~ 26, ruling out an identification of the companion as an OB star, and implying that the system is accreting from a lower mass star, similar to other transient, low-luminosity ULXs[3,30]. This also provides further evidence against the emission being associated with a background active galaxy, which should be substantially brighter.

As the X-ray properties of the source are typical of the class[3,15,17], we confidently extend the identification of Eddington-rate accretion onto a stellar mass black hole in XMMU J004243.6+412519, to the larger population of low-luminosity ULXs (a substantial component of the overall population, with up to 80% having luminosities $1\times10^{39}$ < $L_X$ < $5\times10^{39}$ erg s$^{-1}$)[1]. As a result, this implies that we have a large population of sources from which to develop models for Eddington accretion. In our own Galaxy we are limited to a very small number of such sources, and the large absorbing column through the plane of our Galaxy prevents a clear view of the Eddington-limited accretion flow. However, this is not the case for the vast majority of extragalactic sources, allowing the properties of the infalling matter to be accurately determined. Understanding how the disc and associated outflows behave in this regime will allow us to address outstanding cosmological problems including how outflows from quasars redistributed matter and energy in the early Universe.

Assuming the properties of this source are representative of the larger population, we can consider the possibility of detecting similar events in future. The sensitivity of the VLA would allow us to detect a similarly relativistically-beamed event out to 4 Mpc, or 0.5 Mpc if unbeamed. The X-ray monitoring cadence for M31 has yielded two candidates in ten years, and, although the monitoring was not constant, this is roughly consistent with the observed rate of Galactic BHXRBs with low mass companions whose outbursts reach the Eddington luminosity. It is therefore a realistic prospect (see SI) that future observations of transient ULX systems in nearby galaxies, using sensitive radio telescopes, will permit detailed disc-jet coupling studies and in doing so significantly expand our understanding of Eddington accretion and associated phenomena.

**Acknowledgements:** The authors would like to thank Cath Trott for discussions, and Cody Gough for making his code available.

This work was supported by an STFC standard grant (MM), Netherlands Organization for Scientific Research Vidi Fellowship (SM), European Research Council partial funding (RF) and grant number BMWI/DLR, FKZ 50 OR 1010 (MH).

The National Radio Astronomy Observatory is a facility of the National Science Foundation operated under co-operative agreement by Associated Universities, Inc. We thank the staff of the Mullard Radio Astronomy Observatory for their assistance in the commissioning and operation of AMI, which is supported by Cambridge University and the STFC.

This work is based on observations obtained with *XMM–Newton*, an ESA science mission with instruments and contributions directly funded by ESA Member States and NASA.
This research has also made use of data obtained from NASA's *Swift* and *Chandra* satellites.

**Author Contributions:** MM wrote the manuscript with comments from all authors. JM-J designed and analysed the VLA and VLBA observations. NH-W and AS analysed the AMI-LA observations. J-PM and JM-J carried out the scintillation analysis. SM, RF and MH made significant contributions to the overall science case and manuscript. JC, GB and MG provided support and analysis for the CARMA observations. Remaining authors have either assisted with various aspects of the science case or are contributing members to the M31 group.

**Competing financial interests:** The authors declare no competing financial interests.

**Corresponding author:** All correspondence should be addressed to MM. (m.j.middleton@durham.ac.uk)

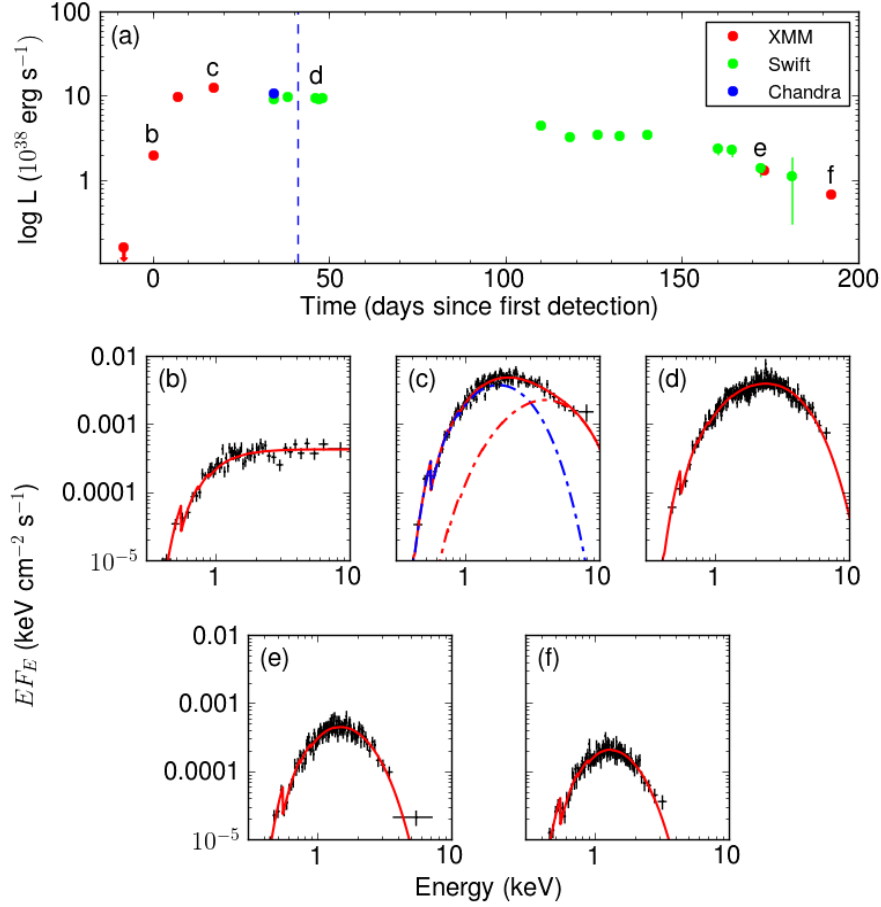

**Figure 1:** Time-evolution of the X-ray luminosity and spectrum. *(a):* Evolution of the source brightness over ~200 days together with the time of the first VLA detection (blue vertical, dashed line). The unabsorbed 0.3-10 keV luminosities (and 90% error bars) are derived from the best-fitting models to the *XMM-Newton* (red), *Chandra* (blue) and *Swift* (green) observations (see Supplementary Table 1 & 2). We also include an upper limit from an *XMM-Newton* observation taken a week before the source brightened. *Bottom panels*: The spectral data (black points with 90% error bars) and best-fitting models (in red, see SI and Supplementary table 2) from which the unabsorbed luminosities are calculated. The first detection prior to the peak brightness shows a spectrum *(b)* that can be fully described by a hard power-law. As the source brightens to a peak luminosity of $1.2 \times 10^{39}$ erg s$^{-1}$, the spectrum *(c)* evolves to one requiring emission from an optically thick accretion disc (dashed blue) and a second, weaker thermal component (dashed red) that may be related to a wind or photosphere. Such spectra are also seen in high mass accretion rate BHXRBs[13] and low-luminosity ULXs[3,15]. As the source dims to $\sim 1 \times 10^{39}$ erg s$^{-1}$ the spectrum *(d)* becomes fully dominated by the disc component. The spectrum remains disc-dominated down to $\sim 7 \times 10^{37}$ erg s$^{-1}$ *(e,f)* which places an upper limit on the mass[10] of $< 17$ M$_\odot$.

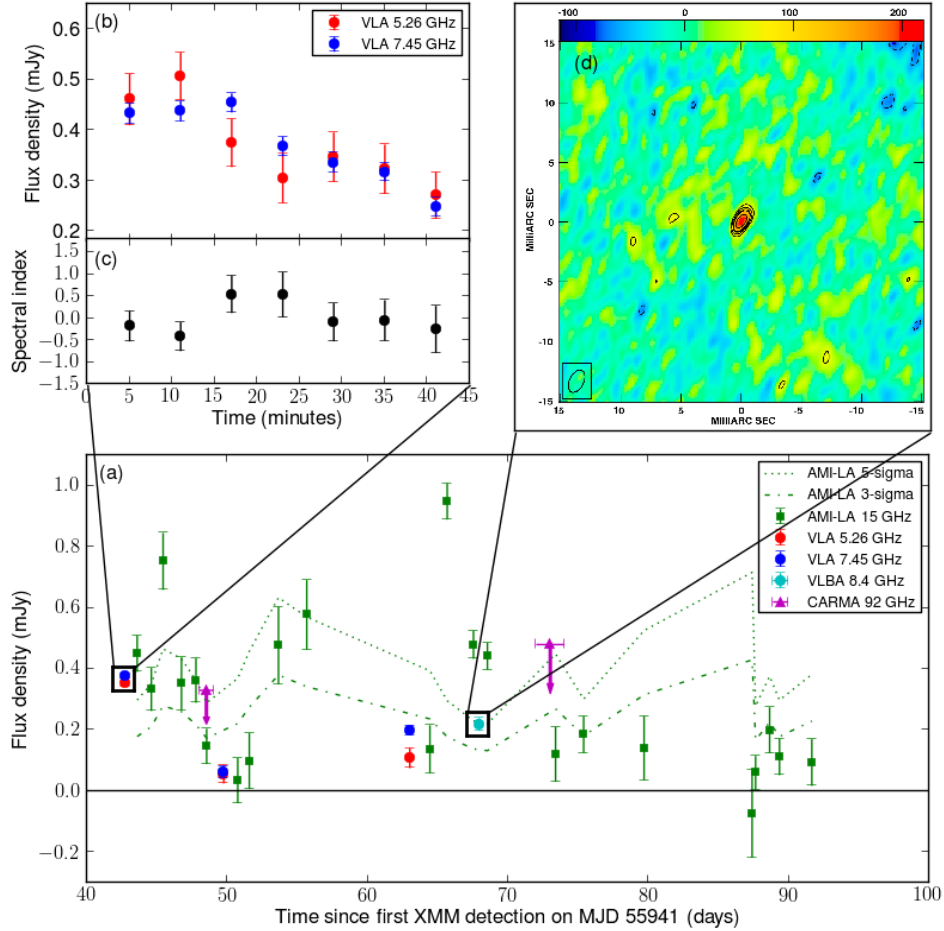

**Figure 2:** Summary of the radio observations. *(a)* Multi-frequency radio light curves of the ULX from our AMI-LA, VLA, VLBA and CARMA data. For the AMI-LA data, we plot the flux density from the pixel value at the source location, owing to dynamic range limitations from the bright, nearby confusing source, 5C 3.111. The resulting high noise level close to the ULX hindered the standard image-plane source fitting methods used for the VLA and VLBA data. To estimate the significance of the AMI points, we have plotted both the 3-sigma and 5-sigma noise levels (dash-dotted and dotted lines), showing both a standard and a (favoured) more conservative detection threshold. *(b)* Radio light curves binned to an interval of 6 minutes (the phase referencing cycle time), showing the short-timescale variability during the first VLA observation. Error bars show the statistical uncertainties at the $1\sigma$ level. There is an additional 1% systematic uncertainty on the overall flux scale. We are unable to conclusively determine whether the observed variability is intrinsic to the source, or whether it arises from scintillation in the interstellar medium. Regardless of the mechanism involved, this constrains the source size to $\leq 6$ microarcseconds. *(c)* Corresponding spectral indices derived from the two VLA observing bands centred at 5.256 and 7.45 GHz. *(d)* VLBA image of the source. Colour scale is in μJy/beam, and the contours are set at levels of $\pm (\sqrt{2})^n$ times the rms noise of 21 μJy/beam, where $n=3,4,5..$ The source is unresolved down to a beam size of $2.0 \times 1.2$ mas$^2$ in position angle -26 degrees east of north. The (J2000) source position, measured relative to J0038+4137 and prior to performing any self-calibration, is RA=00h42m43.674360(9)s, Dec=41d25'18.7037(1)".

**Supplementary Information**

*Karl G. Janksy Very Large Array (VLA)*[31]:
Three VLA observations were taken (see Supplementary Table 3). In all cases, the array was in the relatively compact C-configuration, and we used 3C48 to calibrate both the amplitude scale and the instrumental bandpass, and J0038+4137 to calibrate the antenna amplitude and phase gains. The cycle time between calibration scans was 6 minutes, spending 290 s on target and 70 s on calibrator in each cycle. The target was offset by 7 arcseconds from the field center, to eliminate the possibility of artifacts, generated at the phase center by correlation errors, affecting our measurements.

The original, discovery observation was taken on February 26, 2012 (MJD: 55983) with extremely good weather conditions; clear skies, wind 2.4 - 4.1 km/s, and the API (the atmospheric phase interferometer) showing phase noise of only 0.9-1.9 degrees. This one-hour track was observed in two 1024-MHz basebands centred at 5.256 GHz (at a resolution of $3.7 \times 2.5$ arcsec) and 7.45 GHz (at a resolution of $4.95 \times 3.49$ arcsec), and yielded highly significant detections of $338 \pm 14$ μJy/beam and $371 \pm 8$ μJy/beam respectively (quoted uncertainties at the 1σ level). Images were made using Briggs weighting with a robust factor of 0.5. Two iterations of phase-only self-calibration were performed on short (1-minute) timescales, before longer-timescale amplitude and phase self-calibration was carried out to remove amplitude errors around the two bright, confusing sources in the field.

A second one-hour observation taken on March 4 (MJD: 55990) found the emission to have decayed significantly, giving a weaker detection of $57 \pm 18$ μJy/beam (error of 1σ) when stacking the two C-band basebands. In the 3, 21, 26, and 41 GHz bands, we found 3σ upper limits of 0.17, 0.09, 0.10, and 0.30 mJy/beam respectively. Since the 5.256 and 7.45-GHz detections were only 0.06 mJy/beam, none of these upper limits are constraining. Stacking the data from the 21 and 26-GHz bands, the nominal flux at the source position is -11 μJy/beam with a 1σ noise level of 24 μJy/beam. This high-frequency non-detection suggests a steep rather than inverted spectrum although this is far from conclusive.

A subsequent 30-minute observation taken on March 17 (MJD: 56003) confirmed the radio rebrightening suggested by the AMI-LA monitoring, with observed flux densities of $198 \pm 15$ μJy/beam at 7.45 GHz, and a less significant detection of $108 \pm 33$ μJy/beam at 5.256 GHz . As seen in the first VLA observation, the resulting spectrum is inverted. Importantly the emission was bright enough to trigger the subsequent VLBA observation.

Short timescale variability was only seen in the first of the VLA observations (Figure 2), and we did not see any evidence for significant variability in either the VLBA data or in the final two VLA observations, suggesting that the variability is intrinsic to the source rather than an instrumental artifact. However, to confirm this tentative conclusion, we performed an extensive array of tests to ensure that the reported variability in the first observation was genuine.

1. *Check sources:* A key test to differentiate intrinsic source variability from instrumental or calibration artifacts would be to demonstrate that the observed short-timescale variations were not seen in any other sources in the field. While there are four other unresolved sources in the field that are detected in both the 5.256 and 7.45-GHz images, the rapid decrease of antenna response beyond 1.5 arcmin from the pointing centre means that only the central two sources (within 2 arcmin) can act as reliable check sources, as time-variable pointing errors due to wind loading of the antennas can cause large amplitude variations at larger distances from the pointing centre.

To investigate the relative variability of our target and check sources, we considered the total range of flux densities measured in the light curves, and tested the statistical level at which a constant source is excluded by performing a $\chi^2$ test (summing $\Sigma_i[(S_i-<S>)/\sigma_i]^2$ over all data points). At 7.45 GHz, the total range of flux densities measured for the ULX was a factor of 2.5 greater than that measured for any of the check sources. The $\chi^2$ test showed that the only source for which we could convincingly rule out the null hypothesis of a constant light curve at a statistically significant level (> 95% confidence) was the ULX, providing strong evidence that the variability seen at 7.45 GHz is real. For the 7.45 GHz data, the $\chi^2$ value (for 22 degrees of freedom) was 108.2, implying that the null hypothesis (a constant source) is excluded at greater than 1 part in $10^{19}$. For the 5-GHz data, the $\chi^2$ value was 38.3, implying that probability of the null hypothesis (a constant source) being correct was 3 parts in $10^6$. Thus we are confident that the observed short-timescale variability is real.

2. *Calibration errors*: When binning the data into 6-minute intervals, the amplitude and phase gain solutions on the phase calibrator source were extremely constant in time, so we cannot attribute the observed short-timescale variability to rapidly-changing gain solutions derived using the phase calibrator. Indeed, when blanking every other observation of the phase calibrator, the light curves derived for the target source did not change significantly. Furthermore, the light curves measured for the two independent circular polarisations were identical to within the statistical uncertainties. Additionally, the noise level in the images was constant in time to within 5 μJy/beam, except for the last, shorter time interval, where it rose significantly owing to the reduced exposure time, demonstrating that the short-timescale variability was not caused by overall amplitude fluctuations in the image. Finally, we also checked the effect of carrying out the long-timescale amplitude and phase self-calibration. Since our target source flux density was less than 2% of the total flux density in the image at both frequencies, the self-calibration solutions would be expected to be dominated by the other, constant, sources in the field. A comparison of the light curves of the target source made with and without self-calibration indeed showed that it had no significant impact on the observed variability.

3. *RFI*: As a final test, we investigated the effect of radio frequency interference (RFI) on our light curves. RFI is a narrow-band, time-variable contaminant to the measured visibilities, and could in principle cause spurious variability. While the data were carefully edited to remove any signatures of RFI prior to the external gain calibration, there remains the possibility that spillover of bright RFI from flagged frequency channels into neighbouring channels could have left low-level contamination in the data. To rule out this scenario, we re-ran our analysis omitting any 128-MHz sub-bands that showed any signature of RFI. This left only three of eight sub-bands in the 5.256-GHz baseband, and only two of eight in the 7.45-GHz baseband. While the image noise was higher owing to the reduced bandwidth, the light curves were consistent within uncertainties with those made at full bandwidth, suggesting that RFI was not causing spurious variability.

*Very Long Baseline Array (VLBA)*[34]:

Following the confirmation of the radio rebrightening, we triggered an 8-hour VLBA observation on March 22 2012 (MJD: 56008), using the new wideband backend system to provide a total bandwidth of 256 MHz centred at 8416 MHz in each of two independent circular polarizations, using 2-bit sampling. We used J0038+4137 as a phase reference calibrator (0.83 degrees from the ULX), but further refined the antenna gain solutions by using 5C 3.111 (2 arcminutes from the ULX) as an in-beam calibrator. The self-calibration model included only 5C 3.111. The data were correlated with two separate phase centres, one for the ULX, and one for 5C 3.111. With the available frequency resolution (0.5 MHz channels) and time resolution (2 s), time and bandwidth smearing would have diminished the signal of the ULX in the 5C 3.111 data set to the point where it would not have been detected. Therefore we did not include the ULX in the self-calibration model for 5C 3.111.

Phase solutions on 5C 3.111 were first derived on a 5-minute timescale, and were applied, following which we derived amplitude-and-phase solutions on a 20-minute timescale. Applying these calibration solutions to the ULX made only a 10% difference in the measured flux density of the ULX (increasing it from 0.20 ± 0.02 mJy/beam to 0.22 ± 0.02 mJy/beam). To aid the reader, we provide a VLBA contour plot of 5C 3.111 in Supplementary Figure 1. The contours are at levels of ± ($\sqrt{2}$)$^n$ times the lowest contour level of 25 μJy/beam, where n=3,4,5,... The peak flux density in the image is 11.3 mJy/beam. The selfcal solutions are dominated by the bright central point source (the extension to the south has a maximum brightness of only 0.2 mJy/beam, less than 2% of the peak).

The ULX was detected as an unresolved source at the 11σ level, with a measured brightness of 0.218 ± 0.021 mJy/beam with a beam size of 2.002×1.163 mas. This implies a brightness temperature limit of > 3×10$^6$ K. There was no significant evidence for variability in the VLBA data. The astrometric position for the ULX was derived relative to J0038+4137 (assumed position RA=00h38m24.843587s, Dec = 41d37'06.00021"), and prior to running any self-calibration on 5C 3.111 (i.e. prior to applying the amplitude and phase gains derived using 5C 3.111, since self-calibration can in principle shift positions by a small fraction of a beam). The ULX position was found to be within the *Chandra* X-ray error circle[35] at: RA = 00h42m43.674360(6)s, Dec = 41d25'18.70373(8)". The astrometric errors here are purely statistical, and quoted at the 1σ level. For the calibrator-target throw of 0.83 degrees, we estimate the systematic errors[36] to be < 45 microarcseconds. Adding the systematic and statistical errors linearly gives a final position of: RA = 00h42m43.674360(9)s, Dec = 41d25'18.7037(1)".

An elliptical Gaussian fit to the ULX in the image plane using the AIPS task JMFIT gives a size constraint of < 1.3 × 0.6mas$^2$, with a nominal major axis of 0.8 mas along position angle 138 degrees, and an unresolved minor axis. Fitting in the *uv*-plane using modelfit in Difmap gives a major axis of size 0.9-1.1 mas along position angle -44 to -49 degrees, with the minor axis again unresolved. While both fits suggest a resolved major axis, an unresolved source could not be ruled out, given the uncertainties on the best fitting parameters.

*Arcminute Microkelvin Imager - Large Array (AMI-LA)*[37]:
The AMI-LA has a nominal bandwidth centred at 15 GHz and divided into six usable equal-bandwidth channels of 0.75 GHz spanning 13.5 to 18 GHz. A two hour track by AMI-LA was taken the day following the first observation (February 27, MJD: 55984), yielding a detection of 500 ± 30 μJy/beam at 15 GHz (error of 1σ). Several weeks of monitoring followed this detection, finding a re-brightening to have occurred on March 10 (MJD: 55996), as confirmed by the final VLA detection. Since the resolution of AMI-LA is only 45 arcseconds, the proximity and brightness of 5C 3.111 restricted the achievable dynamic range in the region around the ULX. The ULX flux densities were therefore derived by comparing the pixel value at the VLBA position of the ULX with the rms noise measured in that region of the image.

*Combined Array for Research in Millimeter-wave Astronomy (CARMA)*[38]:
Observations of BHXRB jets have shown the synchrotron emission to extend in an unbroken power-law from the near-infrared down to energies at which self absorption dominates[39]. As CARMA operates at high radio frequencies, this provides an opportunity to constrain the broad-band synchrotron emission from such outflows. The CARMA correlator contains 8 bands, each configured to have 487 MHz bandwidth for optimal continuum sensitivity. The substantial drop in activity detected by AMI-LA in early March was confirmed in the CARMA 3 mm band (centred at 92 GHz), with no detection down to a 3σ upper limit of 0.33 mJy/beam on March 3-4 (MJD: 55989-55990). Subsequent observations on March 27 and 29 (also at 92 GHz, MJD: 56013 and 56015) also showed no detection, to a 3σ upper limit of 0.48 mJy/beam.

*X-ray observations*:
Several observations of the source were taken by *XMM-Newton*[40] (6), *Chandra* ACIS (1)[35] and *Swift*[41] XRT (14), with the observational parameters being described in Supplementary Table 1. We note that, although the full duration of the outburst is unknown, the observed X-ray light curve (Figure 1) appears broadly consistent with those seen during outbursts of Galactic LMXBs[42].

We fit the *XMM-Newton* and *Swift* data using the spectral analysis package, Xspec[43], testing standard BHXRB models for single power-law emission (*po* in Xspec), disc emission (*diskbb* in Xspec), disc emission plus power-law (*diskbb+po*) and disc emission plus thermal Comptonisation (*diskbb+compTT*). In each case we include a neutral absorption column (*tbabs* in Xspec) with the lower limit set to the Galactic line-of-sight column[44] and the best available abundance tables[45]. We find that the first *XMM-Newton* observation when the source was detected (XMM 2) can be fully described by a hard power-law, whereas the outburst observations (XMM 3-4) are best described by emission from a disc with the seed photons Comptonised in a relatively cool population of thermal electrons. The latter is a highly significant improvement over describing the spectral data with *diskbb+po* ($\Delta\chi^2$ of 8 and 59 for each observation, respectively, for an increase of 1 degree of freedom) and is similar to fits obtained for other low-luminosity ULXs and high mass accretion rate BHXRBS[3,13,15,46]. Following the outburst, the *Swift* and *XMM-Newton* data are best described by a single disc component with a temperature very similar to that of high mass accretion rate BHXRBs[9]. We note that we do not detect the presence of any strong edge or line features in any of the X-ray spectra. In addition, we cannot constrain the presence of significant variability in any of the X-ray lightcurves, consistent with observations of other ULXs at similar luminosities[15].

The unabsorbed luminosities from these models (found by including the Xspec component, *cflux*) are given in Supplementary Table 1, and interesting parameters for the best-fitting model in each case (and how each observation corresponds to Figure 1) are given in Supplementary Table 2. For XRT observations 6-14, the data were not of a high enough quality to allow for individual fits and so we fit these simultaneously with XRT 1-5 allowing only the temperature to vary. As a result these luminosities should be treated with caution. However, as the luminosity obtained from an *XMM-Newton* observation (XMM 5) is very close to that obtained by XRT (13) at a very similar epoch, this implies that the quoted luminosities are in fact reasonable estimates.

Following the outburst (XMM 3-4), the X-ray spectrum and luminosity remained remarkably stable on long timescales across five *Swift* observations (XRT 1-5), unlike the lower-frequency radio emission. Of particular importance is the *Swift* observation (XRT 4) that is contemporaneous with the VLA and CARMA non-detections. In addition, a *Chandra* HRC-I observation taken 3 days after the radio rebrightening on March 10, demonstrated that the source had remained bright (with a crude estimated unabsorbed luminosity of $\sim 8\times 10^{38}$ erg s$^{-1}$) in the X-rays (due to the lack of spectral data we do not include this in Figure 1).

By inspecting the behaviour of disc temperature against the unabsorbed luminosity we can see that the emission does not adhere to the L $\propto$ T$^4$ relation seen in BHXRBs at sub-Eddington rates[9] but appears to invert or flatten at the highest luminosities (in XMM(3-4); see Supplementary Figure 2). This trend has been seen in the X-ray spectra of both ULXs[47,48] and in the very high state of some stellar-mass black hole X-ray binaries[49], thus supporting our identification of the source.

*Short-timescale radio variability:*
*1) Brightness Temperature*:
Having established the short-timescale variability as being intrinsic to the source, we can use it to determine the variability brightness temperature and hence constrain the relativistic beaming factor.

Since the brightness temperature is defined as $T_b = S_\nu c^2/(2k_B \nu^2 \Omega)$, we can determine the source size from the variability timescale (the light crossing time), and parameterize the brightness temperature as $T_b = 1\times10^{14}$ $(\Delta S/mJy)$ $(d/kpc)^2$ $(\Delta t/s)^{-2}$ $(\nu/GHz)^{-2}$ K, where $\Delta S$ is the flux density change at frequency $\nu$ in time $\Delta t$, and d is the source distance. For our source, Figure 2 shows a flux density change of 0.21 mJy in 30 minutes at 7.45 GHz, giving a brightness temperature of $T_b = 7\times10^{10}$ K.

While typical Galactic BH XRBs have brightness temperatures determined from resolved VLBI imaging of $T_b < 1\times10^9$ K, there are examples of higher brightness temperatures derived from variability considerations. The well-established 30-minute oscillations in GRS 1915+105[25] (to which we have compared our observations) have brightness temperatures of $T_b = 5\times10^9$ K. However, the BHXRB, V4641 Sgr, has been detected to have a flux density variation of $\Delta S = 24$ mJy at 8.4 GHz in only 4 min[32]. At a source distance of 10 kpc[27], this equates to $T_b = 6\times10^{10}$ K, the same level as we have detected in our ULX. Thus our derived brightness temperature is not unprecedented for a stellar-mass BHXRB.

*2) Detection probability*: Variability brightness temperatures[33] scale as $\delta^3$ (where $\delta=[\gamma(1-\beta \cos i)]^{-1}$ is the Doppler factor). The observed variability brightness temperature for our ULX is $7\times10^{10}$ K, whereas the "standard" brightness temperature for BHXRBs is $1\times10^9$ K, implying a beaming factor of $\delta^3=70$. For an assumed jet speed, we can solve for the required inclination angle to observe the required beaming. For representative jet speeds in the range 0.90-0.98c, the inferred inclination angles are in the range 6-14 degrees. The probability of observing such an event (assuming a uniform distribution of angles to the line of sight) is in the range 0.5-3%. If instead we take the variability brightness temperature of GRS 1915+105 ($5\times10^9$ K) as the reference unbeamed value, the acceptable inclination angles rise to 21-24 degrees, for a probability of 6-8%.

Alternatively, the peak brightness we observed at 780 kpc was 0.4 mJy (corresponding to a few Jy at 10 kpc; a level at which only 5-10 XRBs in our own Galaxy have been detected over the past 20 years). In that time, we have detected outbursts from of order 50 BH and BH candidate XRBs in our own Galaxy, so for a similar population, the probability of detecting such an event is ~10-20%. Since the two estimates are in rough agreement, we estimate a likelihood of detecting the event we observed as a few percent.

Assuming that the VLA can detect such events down to an rms noise level of 5 µJy/beam (feasible in an hour's observing time, assuming 50% overheads), the 3σ detection threshold is then 15 µJy/beam, a factor of 26 below our peak flux. Thus we could detect such an event out to 5 times the distance of M31, i.e. ~4 Mpc. Thus, given the calculated probabilities, we can expect other detections in nearby galaxies, given suitable X-ray/radio monitoring campaigns to detect these sorts of outbursts.

*3) The short-timescale variability and the size of the emitting region*
The observed variability on timescales of several minutes could either be due to intrinsic variations (a variation in the jet speed, electron density, magnetic field, or orientation), or to interstellar scintillation as the signal passes through the turbulent, ionised interstellar medium of our Galaxy[50]. In the case of intrinsic variations, light-crossing time arguments would imply a source size of order 40 light minutes, corresponding to 5 AU, or a few microarcseconds at the distance of M31.

Should the observed variations be intrinsic, and due to a bending of the jet modifying its orientation to the line of sight and hence the Doppler boosting, we can calculate the angle change required. For a jet speed in the range 0.92-0.98c, we require a change in inclination angle of only 3-8 degrees. While this is a plausible scenario to explain the intrinsic variability, the current data do not allow us to determine whether this is the correct explanation. Regardless, the source size constraint from the light-crossing time argument still holds as long as the variability is intrinsic.

Although scintillation is not typically seen in Galactic XRBs, the increased distance to M31 implies an angular size 100 times smaller, such that scintillation might be observed from any relatively compact jet emission. To ascertain whether the observed variability was consistent with what would be expected from scintillation, we used the large bandwidth of the VLA to compute the dynamic spectrum of the source in each of the two basebands, and performed two-dimensional correlation analyses on the dynamic spectra in time-frequency space[51]. The normalised covariance function should show extended features, whose widths in time and frequency give the variability timescale and decorrelation bandwidth of the signal respectively. Should these scale as expected from Kolmogorov turbulence, it would be good evidence that scintillation was responsible for the observed variability. From the variability timescale, we could then derive an upper limit on the source size. While we detected the variability timescale at both frequencies, the decorrelation bandwidth was unconstrained at 7.45 GHz. Thus, while our results were consistent with the expectations of scintillation, there were again too few independent variations in the data to definitively determine the mechanism responsible for the observed short-timescale variability. However, if scintillation was the cause, then for typical scattering screen parameters (a distance $d_s$ of 100 pc and a velocity $v_s$ of 50 km s$^{-1}$), the measured variability timescale ($t=\theta\, d_s/v_s$) implies a source size $\theta$ of 3 microarcseconds.

Thus, regardless of the origin of the short-timescale variability (intrinsic or scintillation), we can place a limit of a few microarcseconds (a few AU at the distance of M31) on the size of the emitting region.

**Supplementary Table 1. Summary of the X-ray observations**

| Instrument (observation) | Observation ID | Date (MJD) | Unabsorbed (0.3-10keV) luminosity ($\times 10^{38}$ erg s$^{-1}$) |
|---|---|---|---|
| XMM (1) | 0674210301 | 2012-01-07 (55933) | < 0.2 |
| XMM (2) | 0674210401 | 2012-01-15 (55941) | 2.0 ± 0.2 |
| XMM (3) | 0674210501 | 2012-01-21 (55947) | 9.8 ± 0.2 |
| XMM (4) | 0674210601 | 2012-01-31 (55957) | 12.6 ± 0.2 |
| XMM(5) | 0700380501 | 2012-07-28 (56136) | 1.3 ± 0.1 |
| XMM(6) | 0700380601 | 2012-08-08 (56147) | 0.7 ± 0.1 |
| ACIS (1) | 13837 | 2012-02-19 (55976) | 10.8 ± 0.4 |
| XRT (1) | 00032286002 | 2012-02-19 (55976) | 9.1 ± 0.5 |
| XRT (2) | 00032286003 | 2012-02-23 (55980) | 9.7 ± 0.6 |
| XRT (3) | 00032286006 | 2012-03-02 (55988) | 9.6 ± 0.4 |
| XRT (4) | 00032286007 | 2012-03-03 (55989) | 9.1 ± 0.4 |
| XRT (5) | 00032286010 | 2012-03-04 (55990) | 9.5 ± 0.5 |
| XRT(6) | 00035336052 | 2012-05-24 (56071) | 4.4 ± 0.2 |
| XRT(7) | 00035336053 | 2012-06-01 (56079) | 3.3 ± 0.2 |
| XRT(8) | 00035336054 | 2012-06-09 (56087) | 3.5 ± 0.3 |
| XRT(9) | 00035336055 | 2012-06-17 (56095) | 3.4 ± 0.3 |
| XRT(10) | 00035336056 | 2012-06-25 (56103) | 3.5 ± 0.4 |
| XRT(11) | 00035336060 | 2012-07-15 (56123) | 2.4 ± 0.4 |
| XRT(12) | 00035336061 | 2012-07-19 (56127) | 2.3 ± 0.4 |
| XRT(13) | 00035336062 | 2012-07-27 (56135) | 1.4 ± 0.2 |
| XRT(14) | 00035336063 | 2012-08-05 (56144) | 1.1 ± 0.8 |

Dates and model unabsorbed luminosities for each X-ray observation of XMMU J004243.6+412519. In addition to the above, a further Chandra HRC-I observation was taken on March 13 which showed that the source was still X-ray bright (at L ~ 8×10$^{38}$ erg s$^{-1}$) when the radio re-brightening occurred (although due to the lack of spectral resolution in this instrument we do not include it in Figure 1).

**Supplementary Table 2. Best-fitting model parameters to the X-ray spectra**

| Instrument (obs) Corresponding panel in Figure 1 | Best-fitting model | | | | Fit quality: $\chi^2$/dof (null hypothesis probability) |
|---|---|---|---|---|---|
| XMM(2) Panel *(b)* | tbabs*po | | | | 161.8/163 (0.51) |
| | nH (×10$^{22}$ cm$^{-2}$) | Index (Γ) | | | |
| | 0.38 ± 0.07 | 2.0 ± 0.2 | | | |
| XMM(3) | tbabs*(diskbb+compTT) | | | | 429.5/457 (0.82) |
| | nH (×10$^{22}$ cm$^{-2}$) | kT$_{disc}$ (keV) | kT$_{compt}$ (keV) | Optical depth (τ) | |
| | 0.45 ± 0.02 | 0.53 $^{+0.16}_{-0.28}$ L = 7.3×10$^{38}$ erg s$^{-1}$ | 1.21 $^{+0.25}_{-0.18}$ | 11.9 $^{+2.1}_{-1.9}$ | |
| XMM(4) Panel *(c)* | tbabs*(diskbb+compTT) | | | | 436.9/423 (0.31) |
| | nH (×10$^{22}$ cm$^{-2}$) | kT$_{disc}$ (keV) | kT$_{compt}$ (keV) | Optical depth (τ) | |
| | 0.44 $^{+0.02}_{-0.05}$ | 0.59 $^{+0.18}_{-0.34}$ L = 10.2×10$^{38}$ erg s$^{-1}$ | 1.27 $^{+0.45}_{-0.33}$ | 11.0 $^{+3.0}_{-2.6}$ | |
| XRT(1-5) Panel *(d)* | tbabs*diskbb | | | | 205.1/203 (0.45) |
| | nH (×10$^{22}$ cm$^{-2}$) | kT$_{disk}$ (keV) | | | |
| | 0.38 ± 0.03 | 0.89 ± 0.02 | | | |
| XMM(5) Panel *(e)* | tbabs*diskbb | | | | 102.4/119 (0.86) |
| | nH (×10$^{22}$ cm$^{-2}$) | kT$_{disc}$ (keV) | | | |
| | 040 $^{+0.03}_{-0.04}$ | 0.45 ± 0.02 | | | |
| XMM(6) Panel *(f)* | tbabs*diskbb | | | | 87.8/112 (0.96) |
| | nH (×10$^{22}$ cm$^{-2}$) | kT$_{disc}$ (keV) | | | |
| | 0.39 ± 0.04 | 0.36 ± 0.02 | | | |

Parameters for each of the best fitting models to the high quality X-ray datasets (uncertainties quoted at the 90% level). In the case of XRT(1-5) we co-add the individual datasets to obtain a well constrained fit. We do not include fits to XRT(6-14) as the data quality is too poor to be fitted in a reliable manner.

**Supplementary Table 3. Summary of VLA observations**

| Date | VLA band (GHz) | Flux density (μJy/beam) |
|---|---|---|
| 2012-02-26 | 5.256 | 338 ± 14 |
| | 7.450 | 371 ± 8 |
| 2012-03-04 | 3.000 | < 168 |
| | 5.256 + 7.450 | 57 ± 18 |
| | 20.800 | < 87 |
| | 25.900 | < 99 |
| | 40.988 | < 312 |
| 2012-03-17 | 5.256 | 108 ± 33 |
| | 7.450 | 198 ± 15 |

Dates, bands and fluxes (with 1σ errors) for the three VLA observations. The observing bandwidth was 1024 MHz except at 41 GHz, where it was 2048 MHz. Both the first and third observations indicate an inverted spectrum whilst the spectrum in the second observation is unconstrained (the quoted upper limits are all at the 3σ level).

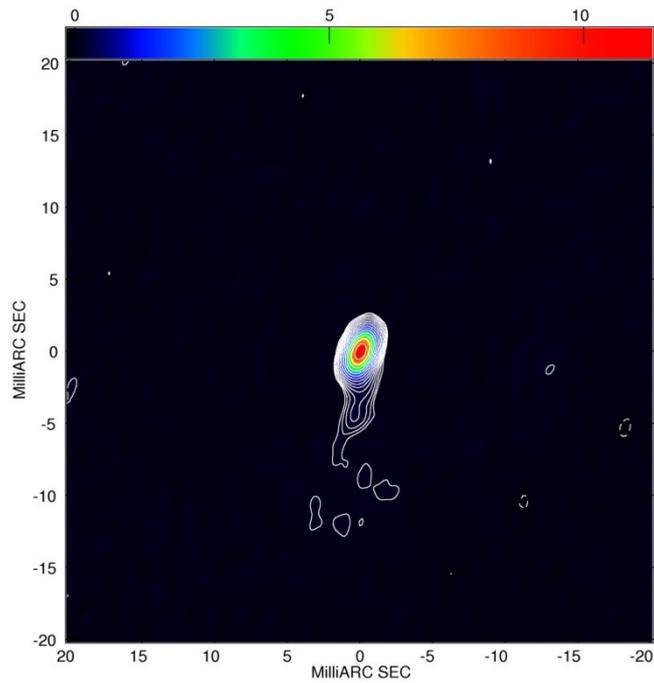

**Supplementary Figure 1:** VLBA image of 5C 3.111. We include this figure to help the reader understand the reliability of the ULX image presented in Figure 2. The contours are at levels of $\pm(\sqrt{2})^n$ times the lowest contour level of 25 µJy/beam, where n=3,4,5,... The peak flux density in the image is 11.3 mJy/beam. The self-calibration solutions are dominated by the bright central point source (the extension to the south has a maximum brightness of only 0.2 mJy/beam, less than 2% of the peak).

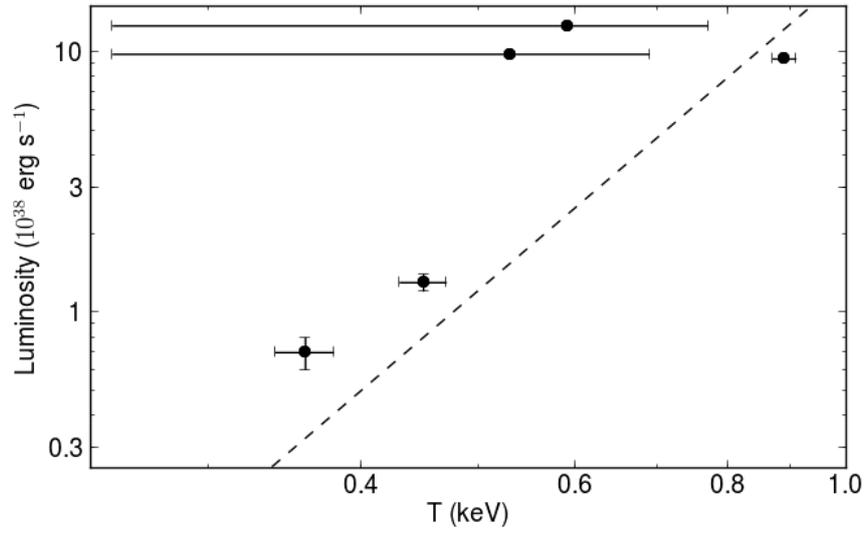

**Supplementary Figure 2:** Log space plot of unabsorbed luminosity versus peak disc temperature (values for which are given in Supplementary Tables 1 and 2) for XMM (3-6) and XRT (1-5). The increase in temperature between the highest luminosity spectrum and the disc dominated spectrum on the decay appears to match the behaviour seen in other low luminosity ULXs[47,48] and in the very high state of some stellar-mass black hole X-ray binaries[49], and is in contrast to that observed in sub-Eddington BHXRBs.